\documentclass[12pt]{article}

\usepackage{amssymb,bbm}
\usepackage{amsfonts}
\usepackage{latexsym}
\usepackage{amsmath}
\usepackage{amsthm}
\usepackage{geometry} 
\usepackage{graphicx}
\usepackage{verbatim}
\usepackage[colorlinks=true]{hyperref} 
\usepackage{bm} 

\newcommand{\II}{{\mathbb I}}

\def\lddots{\mathinner{\mkern1mu\raise1pt\hbox{.}\mkern2mu  
\raise4pt\hbox{.}\mkern2mu\raise7pt\vbox{\kern7pt\hbox{.}}\mkern1mu}}
\makeatletter
\def\numberbysection{\@addtoreset{equation}{section}
 \def\theequation{\thesection.\arabic{equation}}}
\makeatother  
  
\numberbysection

\newcommand{\be}{\begin{eqnarray}}  
\newcommand{\ee}{\end{eqnarray}}

 \title{\bf Reflection matrices from Hadamard-type Temperley-Lieb R-matrices}
\author{ \textsf{Jean Avan$^a$}
\\
\textsf{ Petr Kulish$^c$}
\textsf{ and Genevi\`eve Rollet$^a$}
\\\\
\textit{$^{a}$ Laboratoire de Physique Th\'eorique et Mod\'elisation (CNRS UMR 8089),} \\
\textit{Universit\'e de Cergy-Pontoise, F-95302 Cergy-Pontoise, France} \\
\\
\textit{$^c$
St. Petersburg Department of Steklov Mathematical Institute} \\
\textit{Fontanka 27, 191023, St. Petersburg, Russia} }
\date{}

\footnotetext[1]{Emails: {\tt avan@u-cergy.fr}, 
{\tt kulish@pdmi.ras.ru},
{\tt rollet@u-cergy.fr}}

\begin{document}

\maketitle
\thispagestyle{empty}
\abstract{We classify non-operatorial matrices $K$ solving Skylanin's quantum reflection equation for all $R$-matrices 
obtained from the newly defined general rank-$n$ Hadamard type representations of the Temperley-Lieb algebra $TL_N(\sqrt n)$. 
They are characterized by a universal set of algebraic equations in a specific canonical basis uniquely
defined from the ``Master matrix'' associated to the chosen realization of Temperley-Lieb algebra}

\clearpage 
\newpage

\section{Introduction}
In a previous paper \cite{AKRAnn} we have constructed representations of Temperley-Lieb algebras $TL_N$ \cite{TL}. The $N$
generators were represented by $(n^2 \times n^2)$ dimensional matrices $X_i$, $i = 1 .... N$ identified with a rank $n$ 
matrix $U_{i,i+1}$ acting as an endomorphism on the tensor product of two $n$-dimensional adjacent auxiliary identical vector
spaces $V_i, V_{i+1}$. This construction stemmed from careful consideration of objects proposed in \cite{Chin1,Chin2} in the context
of studies on quantum entanglement.  The matrices $X_i$ thus took the general form 

\be
X_i \equiv \frac{1}{\sqrt n} \sum_{a,b=1}^n e_{ab, (i)} \otimes M^{n_{a}-n_{b}}_{(i+1)}\,,\quad i=1,...,N
\label{form}
\ee
yielding (see e.g. \cite{AKR}) quantum $R$ matrices $R_{i,i+1} = \Pi_{i,i+1} (q' \sqrt n\, \II \otimes \II + T_i)$. $\Pi_{i,j}$ 
generically denotes the permutation operator on tensorized spaces $V_i \otimes V_j$, $\II$ is the $n$-dimensional
identity operator and $q'$ is  the Temperley-Lieb parameter occuring in the normalization of the idempotent relation on $X$:

\be
X_i^2 = -(q' + \frac{1}{q'} ) X_i
\label{normTL}
\ee
Here $q'$ is defined up to an overall inversion by $\sqrt n = q' + \frac{1}{q'}$
The extra factor $\sqrt n$ in the normalization of the term $\II \otimes \II $ is due to our choice
of normalization of $M$ in $X_i$ \ref{form}. 

We have considered the case where the matrices $M$ were diagonalizable as $M = P^{-1} \Lambda P$. The eigenvalues
are thus denoted $\lambda_a,  a=1,...,n$ The classification
achieved in \cite{AKRAnn} relied in particular on the characterization of a certain ``Master matrix'' defined as:

\be
\label{hadspec}
\Omega_{a,b}= \lambda_a^{n_b}\,,\quad a,b=1,...,n
\ee

$\Omega$  must obey the Generalized Hadamard property for a matrix $U$:  $U^{_{-H}} = n\, (U^{-1})^{t}$ where $U^{_{-H}}$ denotes the 
Hadamard inverse:
$(U^{_{-H}})_{i,j} = \frac1{U_{ij}}$. When the matrix elements of $U$ are unimodular this becomes the better known Complex
Hadamard property \cite{Syl,But}.

All consistent matrices $P$ are then obtained by right multiplication of $\Omega^{-1}$ by any matrix $H$ obeying
the generalized Hadamard property.

Following our studies in \cite {AKR} we will develop here the classification of all scalar (i.e. non-operator valued) constant
(i.e. no spectral parameter dependance) matrices $K$,
solutions of the quantum reflection equation \cite{Skl1} associated to these choices of constant $R$-matrix, namely:

\be
R_{12} K_1 R_{12} K_1 = K_1 R_{12} K_1 R_{12}
\label{eqref}
\ee

The purpose of such a classification is to have at our disposal the required ingredients ($R$-matrix and $K$ matrix)
to start building quantum integrable open spin chains with local Hamiltonians: the bulk interaction is controlled
by $R$ and the boundary effects are controlled by $K$. It is of course required to consistently introduce a spectral
parameter dependance in both $R$ and $K$ (a procedure known as Baxterization \cite{Jones}). For a more detailed
discussion of these motivations and an example of such Baxterization for TL representations of $R$ and $K$
matrices see \cite{AKR,Kul03}. Extension of the Baxterization procedure and construction of integrable quantum
spin chains from these elements will be left for further sudies.

We shall derive the explicit equations for matrix elements of $K$ by projecting on suitable generators of $n \times n$ matrices.
The matrix $P$ decouples completely. The Master matrix however plays a crucial role as defining the canonical basis in which
the equations for $K$ take a very simple form, now \textit{independent} of the  Master matrix. This form allows for
complete resolution and parametrization of the $K$ matrices. The situation is significantly different from the classification \cite{AKR}
where the eigenvectors of the $K$ matrix were arbitrary while the eigenvalues depended on the parameters characterizing
the $R$ matrix. Here the eigenvalues will appear as essentially arbitrary while the eigenvectors are at least partially
controlled by the parameters of the $R$ matrix through the Master matrix change of basis.

\section{The fundamental equations}
Let us first of all derive another fully 
algebraic formulation on one single auxiliary space 
for the $K$ matrix equation \ref{eqref} based on the parametrization
\ref{form}, the diagonalization formula $M = P \Lambda P^{-1}$.

\subsection{The algebraic one-space equation}

Parametrizing $K$ as:
\be
K \equiv \sum_{i,j=1}^n K_{ij} e_{ij}
\label{kmat}
\ee

we partially project the reflection equation on the first space generator $e_{il}$ to get a set of algebraic
equations coupling the components of $K$ with powers of the $M$ matrix. Setting $q = \sqrt n q'$ we have:

\be
\sum_{j,k,p=1}^n K_{jk} K_{pl} M^{n_i -n_j +n_k -n_p} + q \sum_{j,s=1}^n K_{js} K_{sl} M^{n_i -n_j }
\ee
\be
= \sum_{j,k,p=1}^n K_{ij} K_{kp} M^{n_j -n_k +n_p -n_l} + q \sum_{j,k=1}^n K_{ij} K_{jk} M^{n_k -n_l }
\label{eqdev}
\ee

The crucial observation is that this equation takes a completely algebraic
form in terms of the matrix $M$. The matrix $P$ of eigenvectors of $M$ thus totally decouples
and only the eigenvalues $\lambda^{(r)}$ and multiplicities $n_i$ characterizing $M$ are relevant
in solving the equations for $K$. The reflection equation \ref{eqdev} thus decouples into $n$ equations
respectively labeled by the eigenvalue label $(r)$ (hereafter set as an index for practical purposes)

\be
\sum_{j,k,p=1}^n K_{jk} K_{pl} \lambda_{(r)}^{n_i -n_j +n_k -n_p} - K_{ij} K_{kp} \lambda_{(r)}^{n_j -n_k +n_p -n_l}
\ee
\be
=  + q \sum_{j,k=1}^n K_{ij} K_{jk} \lambda_{(r)}^{n_k -n_l }- K_{jk} K_{kl} \lambda_{(r)}^{n_i -n_j }
\label{eqdev2}
\ee

We recall that these equations hold for any fixed value of $i,l,r$. Redefining the mute indices $(j,k,p)$ as $(p,j,k)$ 
in the second l.h.s. term and $(j,k)$ as $(k,j)$ in the r.h.s. second term, we can factor out the l.h.s. of \ref{eqdev} as:

\be
(\sum_{j,k=1}^n K_{jk} \lambda_{(r)}^{n_k -n_j })(\sum_{p=1}^n K_{pl} \lambda_{(r)}^{n_i -n_p } - K_{ip} \lambda_{(r)}^{n_p -n_l }
\ee
\be
= q \sum_{j,k=1}^n K_{ik} K_{kj} \lambda_{(r)}^{n_j -n_l }- K_{jk} K_{kl} \lambda_{(r)}^{n_i -n_j }
\label{eqfact}
\ee

Introducing now the matrix $\mu_{(r)}$ defining a rank-one projector for each eigenvalue $\lambda_{(r)}$:
\be
\mu_{(r)} = \sum_{j,k=1}^n \lambda_{(r)}^{n_k -n_j}
\label{mu}
\ee

Equation \ref{eqfact} can now be rewritten as $n$ completely algebraic expressions labeled by $r$,
acting on a single auxiliary vector space, in terms of the matrices $K$ and $\mu_{(r)}$:

\be
(Tr \mu_{(r)} K) (\mu_{(r)} K - K \mu_{(r)}) = q (K^2 \mu_{(r)} -  \mu_{(r)} K^2)
\label{eqalg}
\ee

Complete algebraicity now allows to write \ref{eqalg} in any suitable projection basis of the auxiliary vector space. 
We accordingly introduce the Master basis, characterized by the Master matrix $\Omega$, in which \ref{eqalg} will take a very
simple form when projected on the basis vectors.

\subsection{The Master basis and its component equations}

The Master basis defined by the Master matrix $\Omega$ in \ref{hadspec} consists of the $n$ basis vectors
$v_{(j)}$ with components $\lambda_{(j)}^{n_i}$, $i=1...n$. Acting on any vector $v_{(j)}$ by any matrix $\mu_{(r)}$ yields:

\be
(\mu_{(r)}v_{(j)})^k = \sum_{i=1}^n \lambda_{(r)}^{n_k -n_i} \lambda_{(j)}^{n_i} = \lambda_{(r)}^{n_k} 
\sum_{i=1}^n (\frac{\lambda_{(j)}}{\lambda_{(r)}})^{n_i}
\label{clos1}
\ee

 From the Hadamard generalized condition on $\Omega$ the sum over $i$ is simply given by $n \delta_{rj}$. 
Hence \ref{clos1} simply yields:

\be
\mu_{(r)}v_{(j)} = n \delta_{rj} v_{(j)}
\label{clos2}
\ee

The matrix $\mu_{(r)}$ is thus identified with the diagonal generators $e_{rr}$ in the Master basis. The reexpression
of \ref{eqalg} componentwise in the Master basis considerably simplifies. Let us list the relevant equations.

{\bf 1.} The diagonal contributions to any $r$-labeled equation vanish due to the algebraic forms of commutators of 
$K$ and $K^2$ with the (now) diagonal generator $\mu_{(r)} \equiv e_{rr}$.

{\bf 2.} The off-diagonal contributions to the $r$-th equation separate into two distinct sets respectively obtained from the $rj$ and $jr$
components of \ref{eqalg} in the Master basis:

\be
K_{rr} K_{rj} + q/n (K^2)_{rj} = 0
\label{rj}
\ee

\be
K_{rr} K_{jr} + q/n (K^2)_{jr} = 0
\label{jr}
\ee

One must now separate explicitely diagonal and off-diagonal elements of $K$ when fully expanding $K^2$:

\be
(1 + q/n) K_{rr} K_{rj} + q/n \sum_{s\neq r,j} K_{rs} K_{sj} + q/n K_{rj} K_{jj} =0; j\neq r, j=1...n
\label{rjexp}
\ee
\be
(1 + q/n) K_{rr} K_{jr} + q/n \sum_{s\neq r,j} K_{js} K_{sr} + q/n K_{jj} K_{jr} =0; j\neq r, j=1...n
\label{jrexp}
\ee

Defining now $K \equiv D + K^o $ where $D$ is the diagonal part of $K$ with elements denoted $d_i$,
and $K^o$ its off-diagonal part we rewrite the two sets of $j,r$-labeled conditions as two single matrix equations:

\be
(1 + q/n) D K^o + q/n K^o D + q/n (K^o)^2 = \delta_1
\label{rjalg}
\ee
\be
(1 + q/n) K^o D + q/n D K^o  + q/n (K^o)^2 = \delta_2
\label{jralg}
\ee
where $\delta_1,\delta_2$ are arbitrary diagonal matrices.

This implies in particular that, by substraction of the two equations:
\be
 K^o D - D K^o = \delta_2 -\delta_1 = 0
\label{decoupl}
\ee
since the commutator of $K^o$ with a diagonal matrix $D$ is necessarily non-diagonal. From \ref{decoupl}
it follows immediately that 
\be
d_i \neq d_j \rightarrow K^o_{ij} = 0
\label{class}
\ee

Hence one is naturally lead to discuss the general solutions to \ref{jralg},\ref{rjalg} according to the splitting 
of the diagonal elements $d_i$ of $K$ into subsets of same-valued elements. To each such subset of cardinal
$p$ is then associated a single value of $d_i$ and the corresponding $p \times p$ subblock
of $K^o$; indeed $K^o$ has no non-zero elements outside these subblocks. $(K^o)^2$ is identically structured; hence
one will completely solve \ref{jralg},\ref{rjalg} by considering its reduction to each subset of matrix indices with a 
given value $d$ of the corresponding diagonal element $d_i$ and a given size $p$. 

\section{Resolution of the algebraic conditions}

We now define the reduced equations from \ref{jralg},\ref{rjalg}. We introduce the  $p(d) \times p(d)$ off-diagonal block $K_d^o$ 
associated to a given diagonal element $d$ with a fixed value of $d$. They reduce to a single algebraic equation for $K_d^o$ which reads
\be
(K_d^o)^2 + ( 1 + \frac{2q}{n}) d K_d^o = \delta_p
\label{redalg}
\ee
where $\delta_p$ is an arbitrary $p$-sized diagonal matrix. One sees here that $p(d)$ and $d$ are arbitrary parameters unconstrained
by the equations. In particular, unless $d=0$ (which must be considered separately) every parameter $d$ can be
separately reabsorbed by a rescaling as $K_d^o \rightarrow ( 1 + \frac{2q}{n}) d K_d^o$. The full diagonal sub-block in this last case
therefore exhibits an overall $d$ scaling factor. One can therefore assume that the value of $d$ is fixed to either $0$ or $\frac{n}{n+2q}$. 

To summarize:
The zero-$d$ block $K_0^o$ obeys:
\be
(K_0^o)^2  = \delta_0
\label{redalg0}
\ee
for some diagonal matrix $\delta_0$.

The non-zero-$d$ blocks are all obtained by the above scaling from single universal blocks of arbitrary
size $r$. These universal blocks $K_r$ obey:
\be
(K_r^o)^2 + K_r^o = \delta_r
\label{redalg1}
\ee
for some diagonal matrix $\delta-r$.
 
\subsection{Block with $d=0$}

Let us first consider the case $d=0$. It follows immediately from \ref{redalg0} that $K_0^o$ commutes with the 
diagonal matrix $\delta_0$ hence is again necessarily decomposed into diagonal sub-blocks of size $s\leq p$ with entries in the respective 
cardinal-$s$ sets of indices carried by
each $s$-degenerate diagonal term of $\delta_0$. A reordering of parameters represented by adjoint action
of a permutation matrix of size $p(0)$ allows to rewrite  $K_0^o$ as successive (instead of entangled) diagonal blocks. 
Each such block obeys a simpler algebraic equation:
 \be
(K_0'^o)^2  = \delta_p'. \II
\label{redalg00}
\ee
for arbitrary constants $\delta_p'$
Again two cases must be discussed. 

\subsubsection{The case $\delta_p'$ = 0}

In this case $K_0'^o$ is a nilpotent matrix. All such matrices of size $t$ are represented, up to gauges to be discussed
presently, by two rectangular matrices $A$ and $B$ of size $t \times m$ with $2m \leq t$, as:
\be
K_0'^o = A B^t
\label{nil}
\ee
with the condition that the $m \times m$ matrix $B^t A$ be $0$. $A$ builds a basis of the image vector space of $K_0'^o$
seen as an endomorphism. $B$ builds a basis of the cokernel of $K_0'^o$. The condition $B^t A = 0$ describes the inclusion
of the image in the kernel, a necessary and sufficient condition for nilpotency. Both matrices $A$ and $B$
are defined up to a gauge transformation parametrized by a $m \times m$  
matrix $U$ acting as 
\be
 A \rightarrow AU; B \rightarrow (U^{-1})^tB
\label{gauge2}
\ee

In addition one must impose the vanishing of every diagonal element as
\be
\sum_{a=1}^{m} A_{ia}B_{ia} = 0 \;\; \forall i=1...t
\label{zdiag1}
\ee

The dimension of the moduli space for such matrices with fixed $t,m$ is given by $2mt$ (for the two matrices $A$ and $B$) 
$-m^2$ (gauge freedom \ref{gauge2})
$-m^2$ (condition $B^t A = 0$) $-t$ (condition \ref{zdiag1}) $+1$ (one common equation between $B^t A = 0$ and \ref{zdiag1} due to the
algebraic property Tr$A B^t =$ Tr$B^tA$), yielding an overall dimension $2m(t-m) -t +1$.

More explicit resolution can in general only be achieved on a case-by-case basis. Note however that such blocks, 
coupled as they are to zero diagonal elements of the full $K$ matrix, yield non-invertible $K$ matrices.
 
\subsubsection{The case $\delta_p' \neq 0$}

In this case one rescales every such subblock as  $\bar K_d'^o \equiv \sqrt \delta_d' K_d'^o$ yielding $(\bar K_d'^o)^2  = \II$.
This matrix has zero diagonal terms hence zero trace, its eigenvalues are therefore $\pm 1$ with same degeneracy. Its size $s$
is thus now necessarily even. Any such matrix is again parametrized by two matrices $A$ and $B$ of size $s \times s/2$; respectively
characterizing the eigenspace with eigenvalue $-1$ and $B$ the co-eigenspace of eigenvalue $+1$. In this case it is required
that $B^tA$ be invertible (which is equivalent to requiring zero-intersection between the two eigenspaces)
and one can then always set:
\be
\bar K_d'^o = \II - 2 A (B^t A)^{-1} B^t
\label{ev1}
\ee

A gauge arbitrariness exists also in this parametrization, this time with two independent gauge transformations
acting on $A$ and $B$ interpreted as changes of basis on the two eigenspaces. Namely one can redefine 

\be
 A \rightarrow AU; B \rightarrow B V
\label{gauge3}
\ee
for $U$ and $V$ two arbitrary $s/2 \times s/2$ invertible matrices.

In addition one imposes again the zero-diagonal conditions:
\be
1 - 2 \sum_{a,b=1}^{s/2} A_{ia} (B^t A)^{-1}_{ab}B_{ib} = 0 \;\; \forall i=1...s
\label{zdiag2}
\ee

The moduli space dimension will be discussed together with the general $d \neq 0$ case since the relevant matrices will take exactly
the same form, only the respective degeneracies of eigenvalues will be modified.

A typical example is provided by matrices built with sets of $2 \times 2$ blocks with off-diagonal terms $+1$ . In fact all such matrices
\ref{redalg00} are conjugate to a matrix of this particular type. The requirement that all diagonal terms be zero however
puts constraints on the conjugation matrices which we have not adressed here.

\subsection{Universal $d \neq 0$ blocks}

It follows immediately from \ref{redalg1} that $K_r^o$ commutes with the 
diagonal matrix $\delta_r$ hence it is again necessarily decomposed into diagonal sub-blocks of size $s\leq p$ with entries in the respective 
cardinal-$s$ sets of indices carried by each $s$-degenerate diagonal term of $\delta_r$.  

Again a  reordering of parameters represented by adjoint action
of a permutation matrix of size $r$ allows to rewrite  $K_r^o$ as successive (instead of entangled) diagonal blocks $ K_s^o$
of size $s \times s$. 
Of course the diagonal terms of each subblock remain zero, implying that Tr$K_s^o = 0$. Each such block obeys a simpler 
algebraic equation:
 \be
(K_s^o)^2 + K_s^o = \delta_s \II
\label{redalgr}
\ee
for arbitrary constants $\delta_s$. The minimal polynomial of $K_s^o$ is thus of order at most $2$. If it is
of order $1$ either $K_s^o = 0$ altogether (which is allowed) or $K_s^o = \delta_r' \II$ which is forbidden. Hence
the minimal polynomial is of order $2$ and is exactly given by \ref{redalgr}. 

If $\delta_s = -\frac{1}{4}$ the minimal
polynomial is the exact square $(z + \frac{1}{2})^2$. Hence $K_s^o = -\frac{1}{2} \II + N$ where $N$ is a nilpotent matrix.
Since the trace of $N$ is then necessarily $0$ and the trace of $K_s^o$ is also $0$ this case is eliminated.

Any other value of $\delta_s$ implies that $K_s^o$ is diagonalizable with two distinct eigenvalues $z_1, z_2$. Moreover 
since the trace of $K_s^o$ is $0$ both eigenvalues must be non-zero hence one will never have $\delta_s = 0$

From the form of the minimal polynomial one deduces that $z_1 + z_2 = -1$. In addition the zero-trace condition implies
that $ m' z_1 + (s - m') z_2 =0$ where $m'$ is the multiplicity of $z_1$. The eigenvalues of such a matrix must therefore
take the form
\be
z_1 = \frac{m' -s}{s-2m'}; \;\;\; z_2 = \frac{m'}{s-2m'}
\label{evalues}
\ee
for any integer $m' \leq s$. Note that one cannot have $s-2m' = 0$: the singular behaviour of the eigenvalues reflects
the fact that if the degeneracies are the same the two conditions $z_1 + z_2 = -1$ and $ m' z_1 + (s - m') z_2 =0$ are incompatible.

$K_s^o$ then takes the canonical form (already met with in the previous section) for two-eigenvalue matrices. Introducing
$A$ as the $s \times m'$ rectangular matrix of base vectors with eigenvalue $z_1$, defined up to r.h.s. multiplication
by any $m' \times m'$ invertible matrix $U$ as in \ref{gauge2}; and $B$ as the $s \times m'$ rectangular matrix of base vectors orthogonal to 
the eigenspace with eigenvalue $z_2$ also defined up to r.h.s. multiplication
by any $m' \times m'$ invertible matrix $V$ as in \ref{gauge2} one gets:

\be
K_s^o = z_2 \II + (z_1 -z_2)  A (B^t A)^{-1} B^t
\label{ev22}
\ee

As before invertibility of $B^t A$ is equivalent to the necessary condition of zero-intersection between the two eigenspaces.
One must in addition impose that the diagonal elements of $K_s^o$ vanish, i.e.:

\be
(z_2 -z_1) \sum_{a,b=1}^{m'} A_{ia} (B^t A)^{-1}_{ab}B_{ib} = -z_2 \;\;\forall i=1...s'
\label{zdiag3}
\ee

As a simple example if one eigenvalue (say $z_1$ ) is non degenerate $m' = 1$. $B,A$ are two projective $s$-dimensional
vectors in $CP_{s-1}$ (allowing for overall normalization effects). The eigenvalues are then respectively
 $ z_1 = \frac{1 -s}{s-2}; \;\;\; z_2 = \frac{1}{s-2}$.  The zero-diagonal condition boils down to $b_i a_i = <b|a> \frac{1}{s}\;
\forall i = 1 ... s$. The moduli space here is reduced to one single copy of $CP_{s-1}$.

In general the moduli space has a more complicated structure, but its dimension is easy to compute for  given values
of $m',s$. It yields: $2m's$ (for the matrices $A;B$); $-2m'^2$ (for the gauge arbitrariness $U,V$; $-s$ (for the zero-diagonal
condition) $+1$ (due to the automaticity of the condition Tr$ K_s^o$ given the form \ref{ev22}). It is clearly invariant
under the substitution $m' \rightarrow s-m'$ as should be. It applies to the previous case $d= 0, \delta_p' \neq 0$ with
$m' = s/2$.

\section{Generalization of the two-vector solution}

A generalized version of the Ansatz \ref{form} was proposed in \cite{AKRAnn} in order to properly understand one of the
original examples \cite{Chin1}. It reads:
 \be
X_i \equiv \frac{1}{\sqrt n} \sum_{a,b=1}^n V_a W_b \,\, e_{ab, (i)} \otimes M^{n_{a}-n_{b}}_{(i+1)}\,,\quad i=1,...,N
\label{formVW}
\ee
where $V,W$ are two n-dimensional vectors. This Ansatz also described representations of the Temperley-Lieb algebra
$TL_N(\sqrt n)$. The assumption of diagonalizability for $M$ lead to the $V-W$ Hadamard-type condition for the Master matrix
$\Omega$ in \ref{hadspec}, generally defined for a matrix $U$ as: 
\be
U^{-H} V W U^{t} = tr(W V) \II
\label{VWhad}
\ee
where $V,W$ are recast as diagonal $n \times n$ matrices and $U^{-H}$ is the Hadamard inverse: $(U^{_{-H}})_{i,j} = \frac1{U_{ij}}$.
Plugging now \ref{formVW} into \ref{eqref} we get the same algebraic form as \ref{eqalg} except that :
\be
\mu_{(r)} = \sum_{j,k=1}^n \lambda_{(r)}^{n_k -n_j} V_j W_k
\label{muVW}
\ee
Introducing now the change of basis parametrized by the matrix $\Omega_V$ such that $(\Omega_V)_{ij} \equiv \Omega_{ij}V_j$ 
it is immediate to prove, using \ref{VWhad}, that again in this new basis $\mu_{(r)} = e_{rr}$. The resolution then proceeds
along the original lines of Section 3.

{\bf Acknowledgements}

This work was sponsored by CNRS; Universit\'e de Cergy-Pontoise; Universit\'e de Savoie; and ANR Project DIADEMS (Programme Blanc 
ANR SIMI1 2010-BLAN-0120-02); PPK is partially supported by 
GDRI ``Formation et recherche en physique th\'eorique'' and RFBR grants 11-01-00570-a, 12-01-00207-a.

\end{document}